\documentclass[letter]{aa} 

\usepackage{graphicx}
\usepackage{txfonts}
\newcommand\pmn{PMN~J1603-4904}
\newcommand\fgl{3FGL~J1603.9-4903}

\begin{document}

   \title{Optical-NIR spectroscopy of the puzzling $\gamma$-ray source 3FGL~1603.9-4903/PMN J1603-4904 with X-shooter\thanks{Based on observations collected at the European Organisation for Astronomical Research in the Southern
       Hemisphere, Chile, under program 095.B-0400(A). The raw FITS data files are available in the ESO archive.}}

   \author{P. Goldoni 
          \inst{1}
          \and
          S. Pita\inst{1}
          \and
          C.~Boisson\inst{2}
          \and 
          C.~M\"uller\inst{3,4}
          \and
         T.~Dauser\inst{5}
         \and
         I.Jung\inst{6}
        \and
       F. Krau{\ss} \inst{4,5}
       \and
      J.-P. Lenain\inst{7}
       \and
      H. Sol\inst{2}
}
   \institute{APC, Univ. Paris Diderot, CNRS/IN2P3, CEA/Irfu, Obs. de Paris, Sorbonne Paris Cit\'e
     \and LUTH, Obs. de Paris, CNRS, Universit\'e Paris Diderot, PSL, France
     \and Department of Astrophysics/IMAPP, Radboud University Nijmegen, PO Box 9010, 6500 GL, Nijmegen, The Netherlands
     \and Institut f\"ur Theoretische Physik und Astrophysik, Universit\"at W\"urzburg, Am Hubland, 97074 W\"urzburg, Germany
     \and Dr. Remeis Sternwarte \& ECAP, Universit\"at Erlangen-N\"urnberg, Sternwartstrasse 7, 96049, Bamberg, Germany
     \and ECAP, Universit\"at Erlangen-N\"urnberg, Erwin-Rommel-Str. 1, D-91508 Erlangen, Germany
     \and Sorbonne Universit\'es, UPMC Univ. Paris 06, Univ. Paris Diderot, Sorbonne Paris Cit\'e, CNRS, Laboratoire de Physique Nucl\'eaire et de Haute Energies (LPNHE), 4 Place Jussieu, F-75252, Paris Cedex 5, France
     }
             \offprints{goldoni@apc.in2p3.fr}
             
   \date{}

  \abstract
   {The \textsl{Fermi}/LAT instrument has detected about two thousand extragalactic
     high energy (E $\ge$ 100 MeV) $\gamma$-ray sources. One of the
     brightest is \fgl; it is associated to the radio source \pmn.
     Its nature is not yet clear, it could be either a very peculiar BL Lac or a compact symmetric object
     radio source which are considered as the early stage of a radio galaxy. The latter, if confirmed,
     would be the first detection in $\gamma$-rays for this class of objects.  A redshift z=0.18
     $\pm$ 0.01 has recently been claimed on the basis of the detection of a single X-ray line at 5.44 $\pm$ 0.05 keV
      which has been interpreted as a 6.4 keV (rest frame) fluorescent line.
   }
   {We aim to investigate the nature of \fgl/\pmn~using optical-to-NIR spectroscopy. }
   { We observed \pmn\ with the UV-NIR VLT/X-shooter spectrograph for two hours. We extracted spectra in the
   visible and NIR range that we calibrated in flux and corrected for telluric absorption. We systematically
searched for absorption and emission features. }
 {The source was detected starting from $\sim$ 6300 \AA~down to 24000 \AA~with an
 intensity similar to that of its 2MASS counterpart and a mostly featureless
 spectrum. The continuum lacks absorption features and thus is non-stellar in origin and most
 likely non-thermal. In addition to this spectrum, we detected three emission lines
 that we interpret as the H$\alpha$-[NII] complex, the [SII]$\lambda$,$\lambda$6716,6731 doublet and the [SIII]$\lambda$ 9530 line;
we obtain a redshift estimate of z= 0.2321 $\pm$ 0.0004.
The line ratios suggest that a LINER/Seyfert nucleus powers the emission.
This new redshift measurement implies that the X-ray line previously detected
should be interpreted as a 6.7 keV line which is very peculiar.}
{}

   \keywords{galaxies: active - galaxies: individual: PMNJ1603-4904 -  gamma rays: galaxies - galaxies: distances and redshifts}

   \maketitle
%

\section{Introduction}

  The Large Area Telescope (LAT) onboard the \textsl{Fermi} $\gamma$-ray space telescope has dramatically improved our knowledge of the 
  extragalactic $\gamma$-ray sky at high energies (HE; E > 100 MeV); it has expanded the number of known $\gamma$-ray sources by
an order of magnitude \citep{acero15}. In particular, \textsl{Fermi}/LAT provides important insights into the understanding
of mechanisms at play in blazars which represent the majority ($\sim$85\%) of the $\sim$2000 identified or associated sources of the
3FGL catalog \citep{acero15}. The fact that several of the detected blazars are bright $\gamma$-ray sources has allowed determining
spectral properties and studying temporal variability at different timescales with an accuracy that is unprecedented for this energy
domain. Moreover, the brightest and hardest sources are also detected by the latest generation of very high energy
(VHE; E > 100 GeV) ground based instruments such as H.E.S.S., MAGIC or VERITAS \citep{aha08}, providing stronger
constraints on the energetics and the radiation fields of objects. These
VHE sources will become much more numerous in the next decade with the deployment of the CTA array \citep{act11}.
  
  \textsl{Fermi}/LAT also detected a handful of other extragalactic sources, among them 14 radio galaxies \citep{ack15}.
These sources are thus a minority in the HE sky, but they can provide general insights into the jet structure in AGNs.
Their faintness in the $\gamma$-ray domain prevents a precise spectral and temporal characterization. Therefore, the detection of
bright $\gamma$-ray extragalactic sources that do not belong to the blazar class could allow a significant improvement
in our knowledge of the particle acceleration and radiation mechanisms responsible for $\gamma$-ray emission.
  
  One of the brightest sources detected by \textsl{Fermi}/LAT in the HE $\gamma$-ray sky is \fgl. Its spectrum
 at the highest energies \citep[E > 10 GeV;][]{ack13} shows a photon index of 1.96 $\pm$ 0.14, which makes this source
 a very probable high signal-to-noise detection in the VHE regime. The only radio counterpart which can be found well
 within the 95\% \textsl{Fermi}/LAT error box is \pmn.  On the basis of its $\gamma$-ray properties the source has been classified
  as a low-synchrotron peaked (LSP) BL Lac \citep{ack15}.  Notably, the analysis of the first two years of \textsl{Fermi}-LAT observations
indicated that the source had a very low variability \citep{nol12} while the results of the first four years suggest a high variability
\citep{ack15}. 
 The source lies at $\sim$ 2.5$^{\circ}$ degrees from the Galactic plane in a heavily absorbed region, it has a weak
 optical counterpart \citep{mul14} and a moderately bright NIR counterpart, 2MASSJ16035069-4904054.
 The only spectroscopic observation of the source was performed by \citet{sha13} between $\sim$4000 and $\sim$7000 \AA~with
NTT/EFOSC2 and only 20 minutes of exposure. No emission or absorption lines was detected; this result is
consistent with the LSP classification.

Recently, \citet{mul14}  showed that the source NIR and MIR emission is dominated by thermal radiation that is not
consistent with stellar emission and many times brighter than expected for a typical blazar spectral energy distribution
(SED) in this range. They also found that the emission in the radio band is extended on $\sim$0.01\arcsec\, double-lobed and symmetric.
The symmetry of the radio lobes is difficult to explain if the jet is directed toward the observer as in blazars. Their observations 
suggest two possible scenarios for the source counterpart. One is a very peculiar blazar (most probably a BL Lac)
lacking some of the standard properties of the class such as fast variability.
The second is a radio galaxy whose jet lies in a plane roughly perpendicular to the line of sight.  Because the small spatial
extension of the radio emission is not compatible with a radio galaxy, they suggested that \pmn\ might be a compact
symmetric object (CSO) radio source. This class of sources is defined by the limited spatial extension of its members and is
considered to be the early stage of a radio galaxy \citep{odea98}. While this scenario explains the radio phenomenology well,
no CSO has been firmly detected in $\gamma$-rays as yet, although models have been suggested \citep{sta08}.

 Finally in deep \textsl{XMM} and \textsl{Suzaku} X-ray observations of \pmn\ \citep{mul15}, the source was clearly detected and
modeled with an absorbed power-law spectrum with a spectral index $\Gamma$ = 2.07$^{+0.4}_{-0.1}$ and a Gaussian line
centered at 5.44 $\pm$ 0.05 keV. Interpreting the line as a 6.4 keV fluorescence line,  which is the most prominent line in the X-rays 
because it yields the highest fluorescence, they estimated the redshift at 0.18 $\pm$ 0.01. It is remarkable that, in addition to the Galactic absorption
in the direction of the source, \citep[6.3 $\times$ 10$^{21}$ cm$^{-2}$;][]{kalb05}, an intrinsic absorption three times as strong,
N$_H$ = 2.05$^{+0.14}_{-0.12}$ $\times$ 10$^{22}$ cm$^{-2}$ was detected. 

The nature of \pmn~ is thus not completely determined and information on its properties in the optical/NIR domain
is clearly lacking. We report on the results on optical/NIR spectroscopy with the X-shooter spectrograph
aimed to gather more information on its nature.

For all calculations, we used a cosmology with $\Omega_{\rm M}$ = 0.27, $\Omega_{\rm \Lambda}$ = 0.73 and
H$_0$ = 71 km s$^{-1}$ Mpc$^{-1}$. All wavelengths are measured in vacuum.
  
\begin{figure}
   \centering
   \includegraphics[width=0.9\hsize]{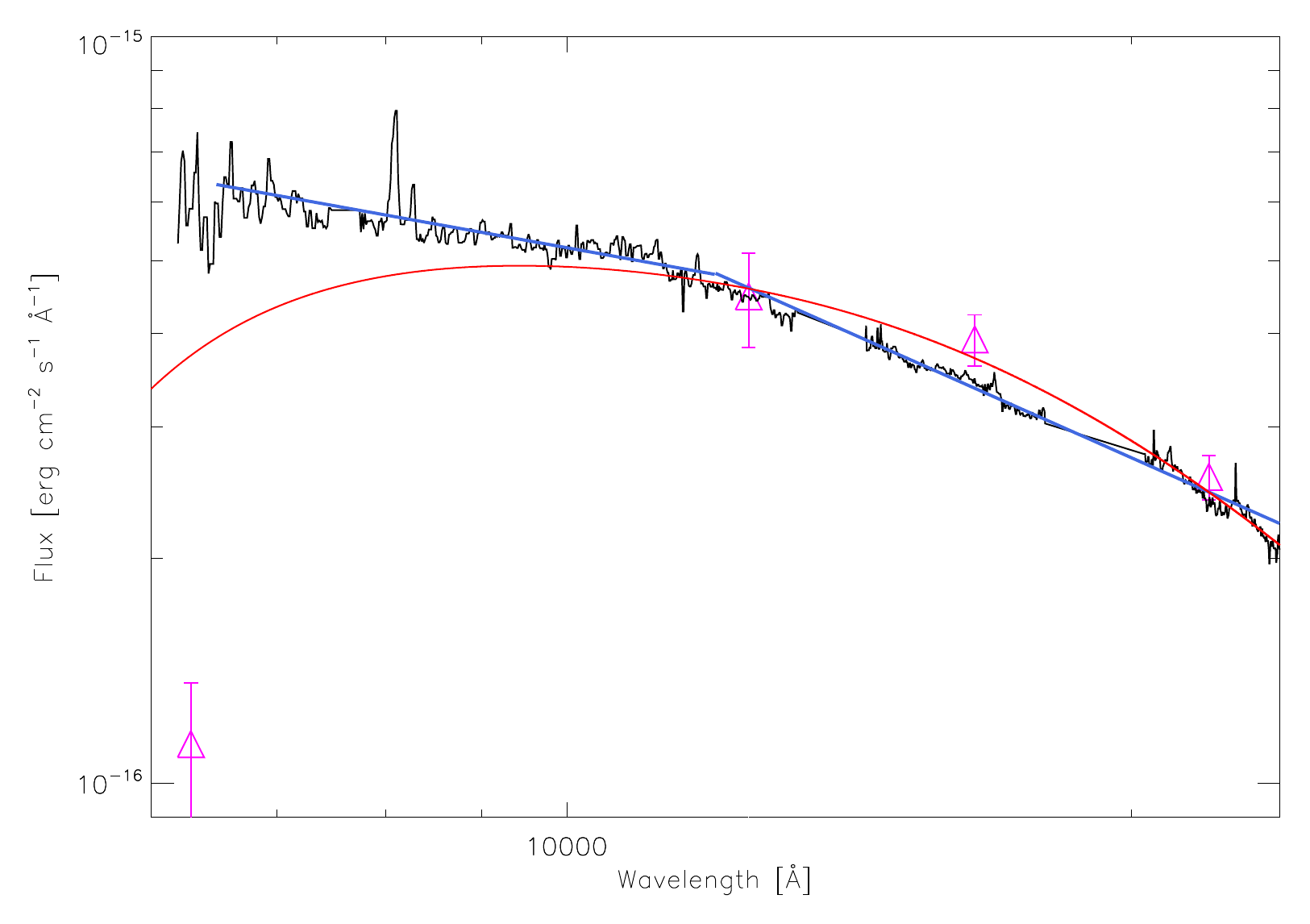}
    \caption{X-shooter dereddened spectrum of PMNJ1603-4904, along with 2MASS photometry and optical photometry \citep{mul14}
   (magenta, triangles with error bars) . The data were dereddened using the parametrization of \cite{fit99} and the E(B-V) of \cite{sch11}.
   We superimposed on the spectra a continuum made by two power laws of index -0.4 and -1.1 in blue and
    a blackbody at T = 1600 K (red). The blackbody cannot reproduce the spectrum at wavelengths shorter than $\sim$ 12000 \AA.}
         \label{figspec}
   \end{figure}


\section{Observations and results}
\label{Observations}

 The X-shooter spectrograph \citep{ver11} is a single-object medium-resolution {\'e}chelle spectrograph
 whose main characteristic is an unprecedented simultaneous wavelength coverage from 3000\,\AA\ to 24000\,\AA. This is obtained
 by splitting the light using dichroics into three arms: UVB ($\lambda$ = 3000 -- 5600\,\AA), VIS ($\lambda$ = 5500 -- 10200\,\AA),
 and NIR ($\lambda$ = 10000 -- 24000\,\AA). Its resolution {$\mathcal R$} is between 3000 and 17000 depending on arm
 and slit width. For these observations, we chose slit widths of 1.3\arcsec, 1.3\arcsec\ and 1.2\arcsec\ for the UVB, VIS,
 and NIR arms, which resulted in {$\mathcal R$} $\sim$ 4000, 6700, and 3900 respectively.

 Two observations were performed, the first on 2015 April 30 and the second about two weeks later on
2015 May 13. The conditions were photometric in the first night and clear in the second. The average airmass
and seeing at zenith of the two nights were 1.27 and 1.14 and 0.76\arcsec\ and 0.57\arcsec\ respectively.
The exposures were taken using the nodding-along-the-slit technique in a standard ABBA sequence with an
offset of  5\arcsec\ between exposures of about 690 seconds each (the exposure time was slightly different for
each arm). Each observation was preceded or followed by an observation of a telluric standard star at similar airmass.
The faint optical emission presents an elliptical shape extended on $\sim$2.5\arcsec\
in the east-west direction and $\sim$1.4\arcsec\ in the north-south direction \citep{mul14}.
Due to nearby sources, the slit was oriented conservatively toward the north, thereby covering only the central
1.2 \arcsec\ (i.e. about half) of the source.

 We processed the spectra using version 2.0 of the X-shooter data reduction pipeline \citep{gol06}.
A description of the detailed reduction and flux calibration process can be found in \citet{pita14}.
The spectra were extracted with a binning of 0.2 \AA~in UVB and VIS data and 0.5 \AA~in the NIR data.
They were then telluric corrected and flux calibrated using the SpeXtool IDL procedure \citep{vac03} and
the SIMBAD magnitudes of the telluric standard stars to set the absolute flux scale. No variability was
detected between the two observations, therefore the two spectra were averaged. 

In the NIR data a bright source with magnitudes compatible with those of the 2MASS counterpart was detected.
The same source was also detected in the VIS arm down to $\sim$ 6300 \AA. Integrating the VIS spectrum
\citep{pita14}, we obtained an equivalent magnitude r' = 22.4 $\pm$0.2. As a result of the slit position (see above),
this value is only applicable to the central half of the source. In previous observations \citep{mul14} the optical counterpart appeared
to have roughly a constant surface brightness and r' $\sim$ 24  for the eastern part (i.e. one third of the source)  .
While the comparison is difficult because the measurements were taken in different parts of the source, this great
discrepancy (of a factor $\sim$ 3 taking into account the size difference) implies that the region of the source
that was observed has varied in the r' band between the two observations. The object was not detected in the UVB data.

We corrected for Galactic absorption using the maps of \citet{sch11} and obtained E(B-V)=2.19 for the direction of \pmn.
This translates into A$_V$=6.8 using the usual relationship R$_V$=A$_V$/E(B-V)=3.1 \citep{rie85}. 
We then dereddened the spectra using the parametrization of \cite{fit99}, the result is shown in Fig. \ref{figspec}.

The detected spectrum is mostly smooth, it does not show narrow lines or the strong absorption lines typical of an elliptical galaxy.
The unabsorbed spectrum is clearly
of non-stellar origin; it can be fit with two power laws (F $\propto$ $\lambda^{\alpha}$) with a break at $\sim$ 12000 \AA. The 
energy indices are $\sim$ -0.4 $\pm$ 0.1 for $\lambda \le$ 12000 \AA~and  -1.1 $\pm$ 0.1 for $\lambda \ge$ 12000 \AA~(see Fig.
 \ref{figspec}). Alternatively the NIR part  of the spectrum may also be fit with a blackbody at T$\sim$ 1600 K as in \cite{mul14}
 but the fit is noticeably worse. From this analysis, it appears that the continuum is nonthermal, at least in the VIS part of the spectrum.
 
\begin{figure}[!tbp]
   \centering
   \begin{center}$
\begin{array}{c}
   \includegraphics[width=0.9\hsize]{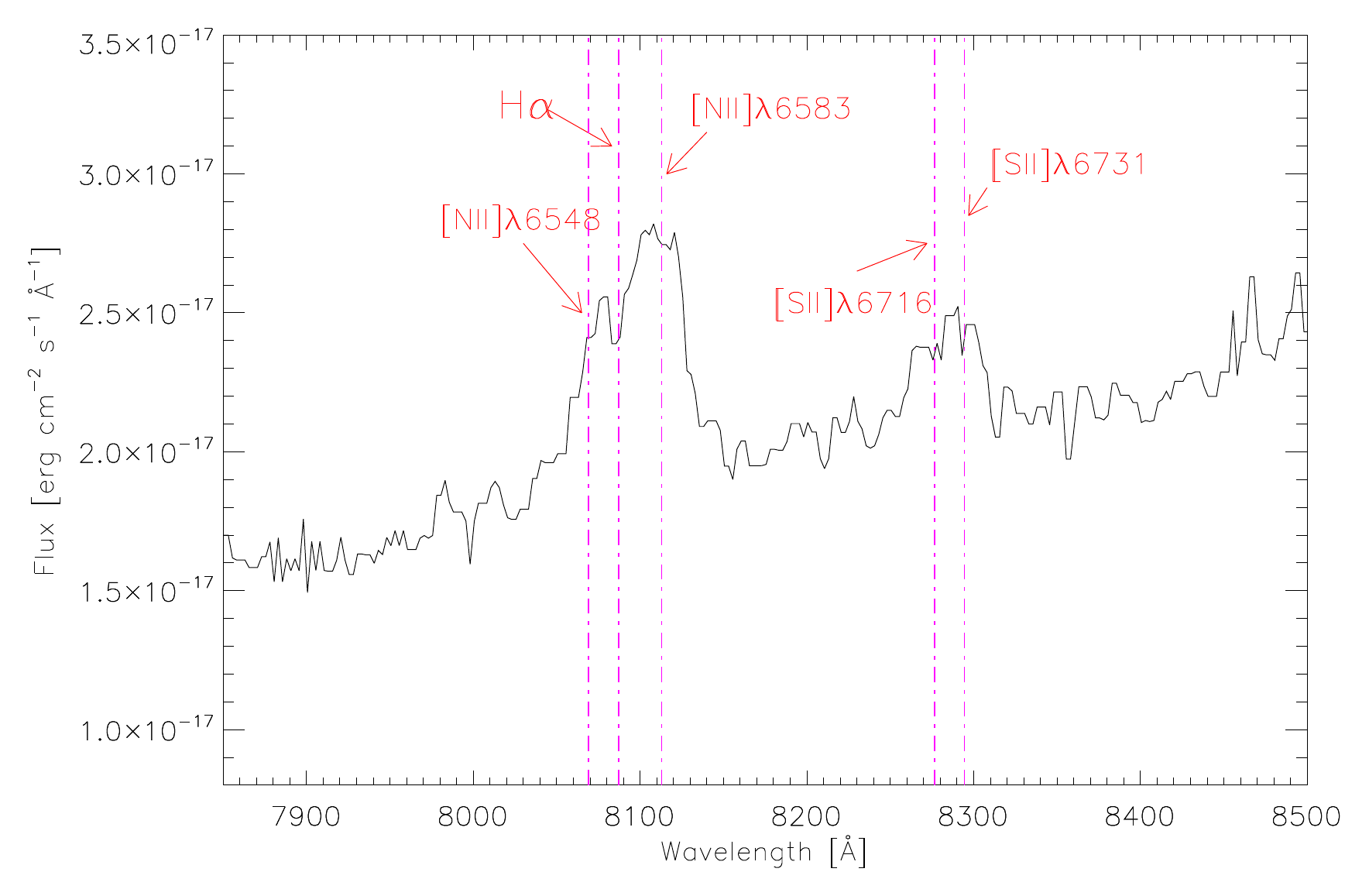} \\

   \end{array}$
      \end{center}

      \caption{Sections of the flux-calibrated and telluric-corrected X-shooter spectrum binned to 2 \AA~bins displaying the
      features described in the text. {\sl Upper panel} H$\alpha$-[NII] complex and [SII]$\lambda$$\lambda$ 6716,6731 doublet. 
      The vertical magenta (dot-dashed) lines show the positions of the feature components (see text).}

         \label{zident}
   \end{figure}

  We did not find any spectral feature consistent with the redshift found by \cite{mul15}. However, we found one bright,
asymmetric, emission feature in the VIS spectrum, centered around 8100 \AA~and a fainter feature centered around 8250 \AA~(Fig \ref{zident}).
In the NIR spectrum we found a much fainter emission line centered around 11700 \AA.
 
  These  three features can be interpreted as the H$\alpha$-[NII] complex (formed by the combination
of H$\alpha$, [NII]$\lambda$ 6583 and [NII]$\lambda$ 6548), the doublet [SII]$\lambda$$\lambda$
6716,6731 and [SIII]$\lambda$ 9530 at a redshift $\sim$ 0.23, respectively. The line [SIII]$\lambda$ 9069, associated
with [SIII]$\lambda$ 9530 and usually weaker by a factor 2.44, is not detected, this is consistent with its
weakness and its position in a wavelength region that is heavily contaminated by telluric absorption.
At this redshift the optical luminosity is 1.0  $\times$ 10$^{45}$ ergs/s and, taking into
account the results of the 3FGL catalog, the $\gamma$-ray luminosity is 8.5 $\times$ 10$^{45}$ ergs/s.
After rebinning the spectra at 2 \AA, we measured the rest-frame equivalent widths by integrating the flux of the three
features taking into account errors in the continuum placement as in \cite{sem92} and obtaining 28.6$\pm$3.3 \AA,
10.2$\pm$1.9 \AA~and 5.7$\pm$1.0 \AA. The first two values are much higher than the usually quoted limit of
5 \AA~for the maximum EW of an emission line in a BL Lac object \citep{urr95}. 

 Fitting a single Gaussian to the 11700 \AA~feature, we estimate a precise redshift of z=0.2321$\pm$0.0004.
The fit of the [SIII]$\lambda$ 9530 results in FWHM$\sim$1300 km/s, but the weakness
of the line and the noisy background render this determination uncertain.
To estimate the parameters of the remaining lines we constrained the velocity widths of the detected
doublets [NII]$\lambda$$\lambda$ 6583,6548 and of [SII]$\lambda$$\lambda$ 6716,6731 to be the same
and the flux ratio [NII]$\lambda$ 6583/[NII]$\lambda$ 6548 equal to 3.
Formally acceptable but weakly constrained fits can be obtained with this configuration; the widths of the lines
are between 1000 and 1800 km/s. With this decomposition we also obtain EW(H$\alpha$) = 9.1$\pm$2.0 \AA~and
EW([NII])=16.8$\pm$3.6 \AA.

Finally we reanalyzed the X-ray spectra presented in \citet{mul15} using the new redshift, the results are essentially
unchanged with spectral index $\sim$ 2.0 and intrinsic N$_H$ $\sim$ 2.3 $\times$ 10$^{22}$ cm$^{-2}$.
In this fit the X-ray line is centered at 6.7 keV, which is very peculiar, but not impossible for this kind of source. 
We also note that a worse fit can be obtained with Galactic absorption N$_H$ = 2$\times$ 10$^{22}$ cm$^{-2}$, but
these column densities are quite rare in this region of the sky, because only 7 \% of the sightlines in a cone of 10
degrees radius around \pmn\ have this absorption in the LAB survey \citep{kalb05}. We therefore consider this possibility unlikely.

\section{Discussion}
\label{Discussion}

 We have detected the optical and NIR emission of the central region of the optical counterpart of
\pmn with X-shooter observations. The continuum spectrum is characterized by non-stellar, most likely nonthermal, emission 
that displays a spectral break around 12000 \AA. In addition of the continuum emission we detected three emission lines consistent with the
H$\alpha$-[NII] complex, the doublet [SII]$\lambda$$\lambda$ 6716,6731 and [SIII]$\lambda$ 9530 at redshift 0.2321
$\pm$ 0.0004.
The equivalent width of the lines is much higher than 5 \AA, the commonly used threshold between
BL Lacs and FSRQs. Blazars can sometimes cross this limit, BL Lac itself has displayed EWs up to
15 \AA\  \citep{verm95}. \citet{ruan14} investigated this phenomenon, and found only six cases of similar transitions in
602 unique repeat pairs of SDSS spectra of 354 blazars. This transition is therefore improbable, but not impossible.
In a more physical classification scheme \citet{ghi11} separated BL Lacs from FSRQs using the Fermi 2FGL
$\gamma$-ray luminosity and spectral index. In this scheme, \pmn\ with 8$\times$ 10$^{45}$ erg/s and 
1.12 $\pm$ 0.025 would marginally be classified as a BL Lac.
(see their Figure 1). However the transition between the two classes is not sharp because sources classified
as BL Lac and FSRQs can be found on both sides of the limit. We conclude that \pmn\ can be tentatively classified
as a blazar, but more observations are needed to determine whether it is a BL Lac or an FSRQ.

We fitted the lines using Gaussian functions for each atomic species. Using this decomposition, we obtained a ratio of
[NII]/H$\alpha$ = 1.3 $\pm$ 0.25. This value is much greater than what can be produced in star-forming regions, for instance $\sim$ 0.4
\citep[see, e.g.][]{bpt81} and can be produced by AGN radiation \citep{kew13}. We then computed the ratio
([SII]$\lambda$ 6716+[SII]$\lambda$ 6731)/H$\alpha$ = 0.59$\pm$0.27 and the ratio ([SIII]$\lambda$ 9069+[SIII]$\lambda$ 9530)/H$\alpha$=
0.47$\pm$0.23. The latter was computed assuming that [SIII]$\lambda$ 9530/[SIII]$\lambda$ 9069 = 2.44. According to \cite{dia85}
(see their Fig 5), these ratios are produced when the ionization is due to the radiation of an AGN. 
  
 The fitted FWHMs lie between $\sim$ 1000 km/s and 1800 km/s and are much greater than the typical FWHMs
\citep[200-300 km/s;][]{Vero97} of emission lines in star-forming regions.  These line widths are intermediate between
the typical widths of lines originating in broad line regions (BLRs) and narrow line regions (NLRs) \citep{Pet06}.
These widths can nonetheless be produced by the combination of several independent narrow
components as occurs in the nearby Seyfert 2 galaxy NGC 1068 \citep{axo98}.  We also searched for evidence
of an additional broad  (FWHM $\ge$ 2000 km/s) H$\alpha$ component and we found that adding this component
does not significantly improve the fit. We conclude that the lines are most likely powered by a LINER or Seyfert nucleus.

 To estimate a limit to the SFR from the absence of narrow lines, we added a narrow (500 km/s) H$\alpha$ line to our fit
 that is representative of the emission from HII regions. Our fit does not reproduce the H$\alpha$-[NII] complex
anymore if the narrow-line H$\alpha$ luminosity becomes greater than 7 $\times$ 10$^{42}$ erg/s
(2$\sigma$ upper limit) implying SFR < 3.8 M$_{\odot}$/yr \citep{ken98}. Therefore the central region of
\pmn\ does not experience a major starburst.
The detection of probably variable non-stellar emission together with moderately bright emission lines suggests
that the object is a radio galaxy powered by a LINER/Seyfert nucleus.  The rather high intrinsic
X-ray absorption column density (2 $\times$ 10$^{22}$ cm$^{-2 }$) is typical of a Seyfert 1.8-1.9
nucleus \citep{ris99} and fits this scenario well. 

 In this case the high-energy (optical to $\gamma$-rays) radiation would be produced through inverse
 Compton upscattering of the photon fields by the particles in the jet \citep{sta08}. The very small
extension in radio ($\sim$0.01\arcsec~are $\sim$40 pc at z=0.2321), typical of CSOs, may be explained if
\pmn\ is a very young radio source whose jet has not had the time to expand. If this were correct, this would be the first
detection of a $\gamma$-ray emitting CSO. Alternatively, \pmn\ could be interpreted as a blazar with a peculiarly
high X-ray absorption and significant iron line emission. In this scenario, the additional X-ray absorption at the galaxy
redshift could be due to a dense interstellar cloud in the path of the jet. A distant foreground cloud causes a similar
column density in the peculiar BL Lac PKS 1413+135 \citep{Perl02}. However in the case of \pmn\, the presence
of the 6.7 keV line suggests that the cloud may be local.

 In both scenarios the presence of a single 6.7 keV line without an associated 6.4 keV line is puzzling, however.
One possible interpretation of this line would be that it is a reflection on a highly ionized thin cloud in the optical
path of the X-ray emission. The cloud would be ionized by interaction of the jet with the ambient medium
 through collisional processes. In this case, the collisionally ionized plasma might even be able to produce the observed 6.7 keV
 emission line without the need of an additional reflection component. Strong photoionization may also produce such
 a line, but a harder and brighter X-ray spectrum would be needed. A more detailed analysis of the origin of this line is
 beyond the scope of this paper.

\section{Conclusions}
\label{Conclusions}

 \pmn~has been suggested to be a very peculiar BL Lac or a young radio galaxy seen edge-on.
 We observed the central region of the optical counterpart with X-shooter. Our main results are: 

$\bullet$ We detected a non-stellar, most likely nonthermal, continuum that varied in the r' band with respect
to earlier observations. In addition we detected three lines at a redshift consistent with 0.2321 $\pm$ 0.0004.

$\bullet$ The EW of the brightest line (the H$\alpha$-[NII] complex) is much greater than 5 \AA, 
but the properties of the source did not allow us to classify it as a BL Lac or as an FSRQ.
The line ratios are consistent with the gas being excited by an AGN nucleus.

$\bullet$ We measured an upper limit to the star formation rate of 3.8 $M_{\odot}$/yr implying that no
strong star formation occurs in the observed region. 

 These results, together with previous multiwavelength observations, led us to consider two
possible scenarios for the nature of the source. In the first scenario, \pmn\ might be a radio galaxy of the CSO
type. This would be the first object of its kind detected in $\gamma$--rays which would permit
a great advancement in our understanding of the mechanism producing $\gamma$-ray
radiation. In the second scenario, the source might be a blazar, whose X-ray emission is obscured through
occultation by a dense intervening and possibly local cloud.
More observations are needed to distinguish between these two scenarios. 
Radio and millimeter observations  aimed at detecting HI 21 cm 
and molecular lines in absorption may also help determine the environment of this object.
New, deeper optical spectroscopy of the totality of the source by changing the position angle
of the slit would be particularly worthwhile. In the central region this would allow checking  for variability
of the continuum and lines. In the eastern and western parts, this would permit searching for the features
of the host galaxy, which will give indications on its orientation with respect to the line of sight and to
the radio source.

\begin{acknowledgements}
     We thank the anonymous referee for a constructive report that helped improve the paper.
We thank C. Martayan, M. Torres,T. Zafar, J. Pritchard and Th. Rivinius for performing the observations
in service mode.
\end{acknowledgements}

\bibliographystyle{aa} 
\bibliography{xsh}        

\end{document}